\documentclass[a4paper,10pt]{article}
\usepackage{graphicx}
\usepackage{comment}
\usepackage{amssymb}
\usepackage{amsmath}
\usepackage{nicefrac}
\usepackage{graphicx}
\usepackage{dcolumn}
\usepackage{bm}
\usepackage{comment}

\begin{document}

\newcommand{\bin}[2]{\left(\begin{array}{c} \!\!#1\!\! \\  \!\!#2\!\! \end{array}\right)}
\newcommand{\hH}{\hat{H}}
\newcommand{\hT}{\hat{T}}
\newcommand{\hX}{\hat{X}}
\newcommand{\hY}{\hat{Y}}
\newcommand{\hZ}{\hat{Z}}
\newcommand{\ha}{\hat{a}}
\newcommand{\hx}{\hat{x}}
\newcommand{\hp}{\hat{p}}

\huge

\begin{center}
Commutation relations of operator monomials
\end{center}

\vspace{0.5cm}

\large

\begin{center}
Jean-Christophe Pain\footnote{jean-christophe.pain@cea.fr}
\end{center}

\normalsize

\begin{center}
\it CEA, DAM, DIF, F-91297 Arpajon, France
\end{center}

\vspace{0.5cm}

\begin{abstract}
In this short paper, the commutator of monomials of operators obeying constant commutation relations is expressed in terms of anti-commutators. The formula involves Bernoulli numbers or Euler polynomials evaluated in zero. The role of Bernoulli numbers in quantum-mechanical identities such as the Baker-Campbell-Hausdorff formula is emphasized and applications connected to ordering problems as well as to Ehrenfest theorem are proposed.
\end{abstract}


\section{Introduction}\label{sec1}

The commutator of two operators $\hX$ and $\hY$ is defined by

\begin{equation}
\left[\hX,\hY\right]=\hX\hY-\hY\hX
\end{equation}

\noindent and the anti-commutator of two operators $\hX$ and $\hY$ as

\begin{equation}
\left\{\hX,\hY\right\}=\hX\hY+\hY\hX.
\end{equation}

\noindent In order to avoid confusion with the Poisson and Lagrange brackets in classical mechanics \cite{GOLDSTEIN80}, the commutator is sometimes written  $\left[\hX,\hY\right]_-$ and the anti-commutator $\left[\hX,\hY\right]_+$. In quantum mechanics, non-commuting operators are very usual, as well as commutators of functions of such operators. For instance, the Baker-Campbell-Hausdorff formula \cite{DYNKIN47,BOSE89}, which links Lie groups to Lie algebras, involves exponential functions. Expanding such functions in Taylor series leads to commutators of monomials, which play a role for instance in the cumulant-expansion techniques \cite{SCHMIDT99,HUGHES09} used in the quantum statistics of damped optical solitons, in the tensorial derivation of oscillator-strength sum rules \cite{KLAR87}, in non-commutative quantum mechanics \cite{SCHOLTZ09,SINHA12} or in the evaluation of averages of products of operators \cite{GINOCCHIO73,OREG90,JONAUSKAS07}. 

In the classical context, operator identities involve the Poisson brackets, while in quantum mechanics the commutators appear instead. This is due to the fact that these identities are based on algebraic properties which are the same for Poisson brackets and commutators, since they are two different realizations of the Lie products. However, anti-commutators are also very important; for instance, the Dirac field Hamiltonian is bounded below only when one uses anti-commutation relations on the creation/annihilation operators instead of commutators \cite{BJORKEN64}. In quantum mechanics, commutators and anti-commutators both arise on an equal footing; it is thus legitimate to ask what analogous identities the anti-commutators do satisfy. Moreover, if some identities exist also for anti-commutators, expressions relating commutators to anti-commutators are lacking and rather difficult to derive.

The importance of (anti-)commutation relations involving monomials is pointed out in section \ref{sec2} through several examples. In section \ref{sec3}, the commutator of monomials of two operators, which commutator is a constant, is expressed as a linear combination of anti-commutators of monomials of the same operators with lower powers. The corresponding formula makes use of Euler polynomials evaluated in zero, which can be related to Bernoulli numbers. The occurence of Bernoulli numbers and anti-commutators in the Baker-Campbell-Hausdorff identity is emphasized in section \ref{sec4} and the role of moments of operators in the time evolution of a quantum system is evoked in section \ref{sec5}. Section \ref{sec6} is the conclusion.

\section{\label{sec2} Quantization, monomials and ordering problems}

\noindent Anti-commutators do not obey the same algebraic properties as commutators. For instance, commutators satisfy the Jacobi identity:

\begin{equation}
\left[\left[\hX,\hY\right],\hZ\right]+\left[\hX,\left[\hY,\hZ\right]\right]+\left[\hY,\left[\hX,\hZ\right]\right]=0,
\end{equation}

\noindent whereas for anti-commutators

\begin{equation}
\left\{\left\{\hX,\hY\right\},\hZ\right\}+\left\{\hX,\left\{\hY,\hZ\right\}\right\}+\left\{\hY,\left\{\hX,\hZ\right\}\right\}\ne 0.
\end{equation}

\noindent Menda\v{s}, Milutinovi\'{c} and Popovi\'{c} discussed in detail the Baker-Hausdorff lemma (also known as the Lie series):

\begin{eqnarray}
\exp\left[\hX\right]\hY\exp\left[-\hX\right]&=&\hY+\left[\hX,\hY\right]+\frac{1}{2}\left[\hX,\left[\hX,\hY\right]\right]\nonumber\\
& &+\frac{1}{6}\left[\hX,\left[\hX,\left[\hX,\hY\right]\right]\right]+\cdots,
\end{eqnarray}

\noindent which is needed for the proof of the Baker-Campbell-Hausdorff theorem and has also many applications \cite{SELSTO07}. The authors found a similar relation for anti-commutators \cite{MENDAS89,MENDAS90,MENDAS10}:

\begin{eqnarray}
\exp\left[\hX\right]\hY\exp\left[\hX\right]&=&\hY+\left\{\hX,\hY\right\}+\frac{1}{2}\left\{\hX,\left\{\hX,\hY\right\}\right\}\nonumber\\
& &+\frac{1}{6}\left\{\hX,\left\{\hX,\left\{\hX,\hY\right\}\right\}\right\}+\cdots,
\end{eqnarray}

\noindent which is more convenient for determining similarity transformations whenever operators $\hX$ and $\hY$ are such that the repeated anti-commutators are simpler to evaluate than the corresponding repeated commutators. Multiplying by $\exp\left[-2\hX\right]$ on the right leads to

\begin{eqnarray}
\exp\left[\hX\right]\hY\exp\left[-\hX\right]&=&\left(\hY+\left\{\hX,\hY\right\}+\frac{1}{2}\left\{\hX,\left\{\hX,\hY\right\}\right\}\right.\nonumber\\
& &+\left.\frac{1}{6}\left\{\hX,\left\{\hX,\left\{\hX,\hY\right\}\right\}\right\}+\cdots\right)\nonumber\\
& &\times\exp\left[-2\hX\right].\nonumber\\
\end{eqnarray}

\noindent In the present work we consider (except in section \ref{sec4}) the case where the commutator is equal to a constant $c$ multiplied by the identity operator $\hat{I}$:

\begin{equation}
\left[\hX,\hY\right]=c~\hat{I},
\end{equation}

\noindent which means that $\left[\hX,\hY\right]$ commutes with $\hX$ and $\hY$. This is the case, for instance, of the canonical relation $\left[\hat{x},\hp_x\right]=i\hbar~\hat{I}$, where $\hat{x}$ and $\hp_x$ are the first coordinates of position and impulsion respectively, or for the bosonic annihilation ($\ha$) and creation ($\ha^{\dag}$) operators which are such that $\left[\ha,\ha^{\dag}\right]=1$. However, relations for commutators obeying different commutation relations can also be obtained (see for instance Ref. \cite{SAU78} for the case $\left[\hX,\hY\right]=\lambda\hY$ where $\lambda$ is a function of $\hX$). 


In the quantization of classical systems one encounters an infinite number of quantum operators corresponding to a particular classical expression. This is due to the numerous possible orderings of the non-commuting quantum operators. Any analytical function of two non-commuting operators $\hX$ and $\hY$ is defined by its power series expansion in terms of these operators:

\begin{equation}
f\left(\hX,\hY\right)=\sum_k\sum_l\sum_m\cdots\sum_nf_{k,l,m,...,n}\hX^k\hY^l\hX^m\cdots\hY^n.
\end{equation}

\noindent A relation between commutators and anti-commutators can help reordering products of operators \cite{MIKHAILOV83,LOHE91,LOHE05}, drawing connections (``intertwinnings'') between normal, anti-normal and Weyl orderings (see also the representation of Glauber-Suddarshan \cite{GLAUBER63,SUDARSHAN63} and Husimi \cite{HUSIMI40} in quantum optics). Born and Jordan \cite{BORN25} proposed a specific ordering of the quantum mechanical momentum and position operators $\hx$ and $\hp_x$, in order to define a hermitian Hamiltonian. they sugested to replace the multinomial quantity $p_x^mx^n$ in classical mechanics by

\begin{equation}
p_x^mx^n\rightarrow \frac{1}{m+1}\sum_{k=0}^m\hp_x^{m-k}~\hx^n~\hp_x^k. 
\end{equation}

\noindent Two years later, Weyl \cite{WEYL27,WEYL50} specified another order, symmetric under the interchange of $\hx$ and $\hp_x$. Bender and Dunne \cite{BENDER89a} defined the multinomial operator $\hT_{m,n}$ as the Weyl-ordered form of the classical function $\hY^m\hX^n$:

\begin{equation}
\hT_{m,n}=\frac{1}{2^n}\sum_{k=0}^n\bin{n}{k}\hX^k\hY^m\hX^{n-k}.
\end{equation}

\noindent The polynomial $\hT_{m,n}$ can be expressed as the totally symmetrized form containing $m$ factors of $\hX$ and $n$ factors of $\hY$, normalized by the number of terms in the expression. One has, for instance 

\begin{eqnarray}
\hT_{1,1}&=&\frac{1}{2}\left(\hX\hY+\hY\hX\right)=\frac{1}{2}\left\{\hX,\hY\right\}\nonumber\\
\hT_{1,2}&=&\frac{1}{3}\left(\hX^2\hY+\hX\hY\hX+\hY\hX^2\right)\nonumber\\
\hT_{2,2}&=&\frac{1}{6}\left(\hX^2\hY^2+\hY^2\hX^2+\hX\hY\hX\hY+\hY\hX\hY\hX+\hX\hY^2\hX+\hY\hX^2\hY\right)\nonumber\\
\hT_{0,4}&=&\hX^4.
\end{eqnarray}

\noindent The following relation \cite{BENDER89b}

\begin{eqnarray}
\hT_{m,n}\hT_{r,s}&=&\sum_{j=0}^{\infty}\frac{\left(\frac{i}{2}\right)^j}{j!}\sum_{k=0}^j(-1)^{j-k}k!^2(j-k)!^2\bin{j}{k}\nonumber\\
& &\times\bin{m}{j-k}\bin{n}{k}\bin{r}{k}\bin{s}{j-k}\nonumber\\
& &\times\hT_{m+r-j,n+s-j}
\end{eqnarray}

\noindent enables one to obtain the expressions of the commutator

\begin{eqnarray}
\left[\hT_{m,n},\hT_{r,s}\right]&=&2\sum_{j=0}^{\infty}\frac{\left(\frac{i}{2}\right)^{2j+1}}{(2j+1)!}\sum_{k=0}^{2j+1}(-1)^kk!^2(2j+1-k)!^2\bin{2j+1}{k}\nonumber\\
& &\times\bin{m}{k}\bin{n}{2j+1-k}\bin{r}{2j+1-k}\bin{s}{k}\nonumber\\
& &\times\hT_{m+r-2j-1,n+s-2j-1}\nonumber\\
\end{eqnarray}

\noindent and the anti-commutator

\begin{eqnarray}
\left\{\hT_{m,n},\hT_{r,s}\right\}&=&2\sum_{j=0}^{\infty}\frac{\left(\frac{i}{2}\right)^{2j}}{(2j)!}\sum_{k=0}^{2j}(-1)^kk!^2(2j-k)!^2\bin{2j}{k}\nonumber\\
& &\times\bin{m}{k}\bin{n}{2j-k}\bin{r}{2j-k}\bin{s}{k}\nonumber\\
& &\hT_{m+r-2j,n+s-2j}.
\end{eqnarray}

\noindent In addition, particular Bender-Dunne polynomials can be expressed as anti-commutators. For instance, one has

\begin{equation}
\hT_{m,m+k}=\frac{(2m+k)!m!}{2(2m)!(m+k)!}\left\{\hT_{m,m},\hX^k\right\}
\end{equation}

\noindent and

\begin{equation}
\hT_{m+k,m}=\frac{(2m+k)!m!}{2(2m)!(m+k)!}\left\{\hT_{m,m},\hY^k\right\}.
\end{equation}

\noindent Cahill and Glauber \cite{CAHILL69a,CAHILL69b,FUJII04,VARGAS06} introduced the concept of ``$s$-ordered displacement operator'' by

\begin{equation}
D(\alpha,s)=e^{\alpha \ha^{\dag}-\alpha^*\ha}e^{s|\alpha|^2/2}
\end{equation}

\noindent where $\alpha$ is a complex number and $\alpha^*$ its conjugate. By the Baker-Campbell-Hausdorff formula for the three discrete values of $s$=1, 0 and -1, the operator $D(\alpha,s)$ can be written as

\vspace{2mm}

\noindent $\rightarrow$ normal order

\begin{equation}
D(\alpha,1)=e^{\alpha \ha^{\dag}}e^{-\alpha^*\ha}
\end{equation}

\noindent $\rightarrow$ Weyl order

\begin{equation}
D(\alpha,0)=e^{\alpha \ha^{\dag}-\alpha^*\ha}
\end{equation}

\noindent $\rightarrow$ and anti-normal order

\begin{equation}
D(\alpha,-1)=e^{-\alpha^*\ha}e^{\alpha \ha^{\dag}}.
\end{equation}

\noindent The authors defined the $s$-ordered product $\vdots\left(\ha^{\dag}\right)^n\ha^m\vdots_{(s)}$ as
 
\begin{equation}
\vdots\left(\ha^{\dag}\right)^n\ha^m\vdots_{(s)}\equiv\left.\frac{\partial^{n+m}D\left(\alpha,s\right)}{\partial\alpha^n\partial\left(-\alpha^*\right)^m}\right|_{\alpha=0}.
\end{equation}

\noindent The intertwinning formula is 

\begin{equation}
D\left(\alpha,s\right)=e^{(s-t)|\alpha|^2/2}D\left(\alpha,t\right)
\end{equation}

\noindent and one can write

\begin{eqnarray}\label{ord}
\vdots\left(\ha^{\dag}\right)^n\ha^m\vdots_{(s)}&=&\sum_{k=0}^{\min(n,m)}k!\bin{n}{k}\bin{m}{k}\left(\frac{t-s}{2}\right)^k\nonumber\\
& &\times\vdots\left(\ha^{\dag}\right)^{n-k}\ha^{m-k}\vdots_{(t)}.
\end{eqnarray}

In the second-quantization theory of atomic spectroscopy, expressing $\left[\ha^n,\left(\ha^{\dag}\right)^m\right]$ in terms of $\left\{\ha^i,\left(\ha^{\dag}\right)^j\right\}$ can be helpful \cite{JUDD67}. Indeed, in order to know the dependence of the operators with respect to the number of particles, a matrix element is written as a product of annihilation and creation operators, and the creation operators must be moved to the left (the annihilation operators being moved to the right) with the help of anti-commutation relations.

\section{\label{sec3} Expression of a commutator of monomials in terms of anti-commutators}

The commutator of functions of operators with constant commutation relations reads

\begin{equation}
\left[f\left(\hX\right),g\left(\hY\right)\right]=-\sum_{k=1}^{\infty}\frac{(-c)^k}{k!}f^{(k)}\left(\hX\right)g^{(k)}\left(\hY\right),
\end{equation}

\noindent where $f^{(k)}$ and $g^{(k)}$ represent the $k^{th}$ derivatives of $f$ and $g$ respectively. In the case where $f\left(\hX\right)=\hX^n$ and $g\left(\hY\right)=\hY^m$, one has \cite{WILCOX67}:

\begin{eqnarray}
\left[\hX^n,\hY^m\right]&=&-\sum_{k=1}^{\mathrm{min}(n,m)}\frac{(-c)^kn!m!}{k!(n-k)!(m-k)!}\hX^{n-k}\hY^{m-k}\nonumber\\
&=&-\sum_{k=1}^{\mathrm{min}(n,m)}(-c)^kk!\bin{n}{k}\bin{m}{k}\hX^{n-k}\hY^{m-k}.\nonumber\\
& &
\end{eqnarray}

\noindent In the same way, since $\left[\hY,\hX\right]=-c~\hat{I}$ and $\left[\hX^n,\hY^m\right]=-\left[\hY^m,\hX^n\right]$, one can write

\begin{eqnarray}\label{int}
\left[\hX^n,\hY^m\right]&=&\sum_{k=1}^{\mathrm{min}(n,m)}\frac{c^kn!m!}{k!(n-k)!(m-k)!}\hY^{m-k}\hX^{n-k}\nonumber\\
&=&\sum_{k=1}^{\mathrm{min}(n,m)}c^kk!\bin{n}{k}\bin{m}{k}\hY^{m-k}\hX^{n-k}.
\end{eqnarray}

\noindent Let us look for an expression of the kind

\begin{equation}\label{init}
\left[\hX^n,\hY^m\right]=\sum_{k=1}^{\mathrm{min}(n,m)}c^kk!\bin{n}{k}\bin{m}{k}v_k\left\{\hX^{n-k},\hY^{m-k}\right\},
\end{equation}

\noindent where $v_k$ is to be determined. Using Eq. (\ref{int}), one has:

\begin{eqnarray}\label{use}
\left[\hX^{n-k},\hY^{m-k}\right]&=&\sum_{\ell=1}^{\mathrm{min}(n-k,m-k)}c^{\ell}\ell!\bin{n-k}{\ell}\bin{m-k}{\ell}\nonumber\\
& &\times\hY^{m-k-\ell}\hX^{n-k-\ell},\nonumber\\
\end{eqnarray}

\noindent which can be put in the form

\begin{eqnarray}
\hX^{n-k}\hY^{m-k}&=&\sum_{\ell=1}^{\mathrm{min}(n-k,m-k)}c^{\ell}\ell!\bin{n-k}{\ell}\bin{m-k}{\ell}\hY^{m-k-\ell}\hX^{n-k-\ell}\nonumber\\
& &+\hY^{m-k}\hX^{n-k}\nonumber\\
&=&\sum_{\ell=0}^{\mathrm{min}(n-k,m-k)}c^{\ell}\ell!\bin{n-k}{\ell}\bin{m-k}{\ell}\hY^{m-k-\ell}\hX^{n-k-\ell}.\nonumber\\
& &
\end{eqnarray}

\noindent Inserting the latter expression into Eq. (\ref{init}) yields

\begin{eqnarray}
\left[\hX^n,\hY^m\right]
&=&\sum_{k=1}^{\mathrm{min}(n,m)}\sum_{\ell=0}^{\mathrm{min}(n-k,m-k)}c^{k+\ell}\frac{n!m!}{k!\ell!(n-k-\ell)!(m-k-\ell)!}\nonumber\\
& &\times v_k~\hY^{m-k-\ell}\hX^{n-k-\ell}\nonumber\\
& &+\sum_{k=1}^{\mathrm{min}(n,m)}c^kk!\bin{n}{k}\bin{m}{k}v_k\hY^{m-k}\hX^{n-k},
\end{eqnarray}

\noindent which can be rewritten

\begin{eqnarray}\label{befper}
\left[\hX^n,\hY^m\right]&=&\sum_{k=1}^{\mathrm{min}(n,m)}\sum_{\ell=k}^{\mathrm{min}(n,m)}c^{\ell}\frac{n!m!}{k!(\ell-k)!(n-\ell)!(m-\ell)!}\nonumber\\
& &\times v_k~\hY^{m-\ell}\hX^{n-\ell}\nonumber\\
& &+\sum_{k=1}^{\mathrm{min}(n,m)}c^kk!\bin{n}{k}\bin{m}{k}v_k\hY^{m-k}\hX^{n-k}.
\end{eqnarray}

\noindent After permutation of the roles of indices $k$ and $\ell$, Eq. (\ref{befper}) becomes 

\begin{eqnarray}
\left[\hX^n,\hY^m\right]&=&\sum_{k=1}^{\mathrm{min}(n,m)}\sum_{\ell=1}^kc^{k}\frac{n!m!}{\ell!(k-\ell)!(n-k)!(m-k)!}v_{\ell}~\hY^{m-k}\hX^{n-k}\nonumber\\
& &+\sum_{k=1}^{\mathrm{min}(n,m)}c^kk!\bin{n}{k}\bin{m}{k}v_k\hY^{m-k}\hX^{n-k},
\end{eqnarray}

\noindent which is equivalent to

\begin{eqnarray}
\left[\hX^n,\hY^m\right]&=&\sum_{k=1}^{\mathrm{min}(n,m)}c^kk!\bin{n}{k}\bin{m}{k}\left[\sum_{\ell=1}^k\bin{k}{\ell}v_{\ell}\right]\hY^{m-k}\hX^{n-k}\nonumber\\
& &+\sum_{k=1}^{\mathrm{min}(n,m)}c^kk!\bin{n}{k}\bin{m}{k}v_k\hY^{m-k}\hX^{n-k}.
\end{eqnarray}

\noindent Gathering the two terms of the right-hand side, one obtains 

\begin{eqnarray}\label{forid}
\left[\hX^n,\hY^m\right]&=&\sum_{k=1}^{\mathrm{min}(n,m)}c^kk!\bin{n}{k}\bin{m}{k}\left[v_k+\sum_{\ell=1}^{k}\bin{k}{\ell}v_{\ell}\right]\nonumber\\
& &\times\hY^{m-k}\hX^{n-k}.\nonumber\\
\end{eqnarray}

\noindent Identifying Eqs. (\ref{int}) and (\ref{forid}) yields

\begin{equation}\label{mt2}
v_k+\sum_{\ell=1}^{k}\bin{k}{\ell}v_{\ell}=1.
\end{equation}

\noindent In the following, we show that it is possible to find a solution in terms of Euler polynomials evaluated in zero, which in turn can be expressed by Bernoulli numbers. The Euler polynomial $E_{\ell}(x)$ is defined by

\begin{equation}
E_{\ell}(x)=\left.\frac{\partial^{\ell}}{\partial t^{\ell}}\left(\frac{2~\exp\left(xt\right)}{\exp(t)+1}\right)\right|_{t=0}
\end{equation}

\noindent and obeys the relations (see Refs. \cite{ABRAMOWITZ65,DLMF})

\begin{equation}\label{rel1e}
E_{\ell}(x+1)+E_{\ell}(x)=2x^{\ell}
\end{equation}

\noindent and

\begin{equation}\label{rel2e}
E_k(x+h)=\sum_{\ell=0}^k\bin{k}{\ell}E_\ell(x)h^{k-\ell}.
\end{equation}

\noindent Inserting the left-hand side of Eq. (\ref{rel2e}) for $h=1$ into Eq. (\ref{rel1e}) yields

\begin{equation}
\sum_{\ell=0}^k\bin{k}{\ell}E_{\ell}(x)+E_k(x)=2x^k.
\end{equation}

\noindent Evaluating the latter expression for $x=0$ leads to

\begin{equation}\label{teme}
E_k(0)+\sum_{\ell=0}^k\bin{k}{\ell}E_{\ell}(0)=0,
\end{equation}

\noindent where $E_{\ell}(0)$ can be evaluated by the explicit expression \cite{EULER}:

\begin{equation}
E_{\ell}(0)=2^{-\ell}\sum_{j=1}^{\ell}\left[(-1)^{j+\ell+1}j^{\ell}\sum_{p=0}^{\ell-j}\bin{\ell+1}{p}\right].
\end{equation}

\noindent Therefore, since $E_0(x)=1$, Eq. (\ref{teme}) reveals that Eq. (\ref{mt2}) has the solution $v_{\ell}=-E_{\ell}(0)$ and the final result is

\begin{equation}\label{finee}
\left[\hX^n,\hY^m\right]=-\sum_{k=1}^{\mathrm{min}(n,m)}c^kk!\bin{n}{k}\bin{m}{k}E_k(0)\left\{\hX^{n-k},\hY^{m-k}\right\}.
\end{equation}

\noindent The solution $v_k=-E_k(0)$ is the unique solution of the system (see appendix). It is interesting to point out that the left hand side of Eq. (\ref{finee}) is a commutator, and therefore anti-symmetric with respect to the interchange of $\hX$ and $\hY$, while its right hand side involves an anti-commutator, which is {\it de facto} symmetric. But it should be noted that if $\hX$ and $\hY$ are permutated, $c$ must be replaced by $-c$. Equation (\ref{finee}) is not a recurrence relation, which should relate the commutator (anti-commutator) of higher powers of $\hX$ and $\hY$ to the commutators (anti-commutators) of the lower powers of these operators. Indeed, the idea behind a recurrence relation is the possibility of iteration; a true recurrence relation would be an expression relating the commutator (anti-commutator) of higher powers of $\hX$ and $\hY$ to the commutators (anti-commutators) of the lower powers of these operators. 

\section{\label{sec4} Role of Bernoulli numbers in quantum-mechanical identities}

Since 

\begin{equation}
E_k(0)=-2\frac{\left(2^{k+1}-1\right)}{k+1}B_{k+1},
\end{equation}

\noindent equation (\ref{finee}) can also be expressed in terms of Bernoulli numbers:

\begin{eqnarray}\label{finee2}
\left[\hX^n,\hY^m\right]&=&2\sum_{k=1}^{\mathrm{min}(n,m)}c^kk!\bin{n}{k}\bin{m}{k}\frac{(2^{k+1}-1)}{k+1}B_{k+1}\nonumber\\
& &\times\left\{\hX^{n-k},\hY^{m-k}\right\}.
\end{eqnarray}

\noindent The Bernoulli polynomial $B_{\ell}(x)$ of order $\ell$ can be obtained by successive derivation of a generating function

\begin{equation}
B_{\ell}(x)=\left.\frac{\partial^{\ell}}{\partial t^{\ell}}\left(\frac{t~\exp\left(xt\right)}{\exp(t)-1}\right)\right|_{t=0}
\end{equation}

\noindent where $B_{\ell}(0)$ is the Bernoulli number of order $\ell$, denoted $B_{\ell}$, which is non-zero only if $\ell$ is even. It is worth mentioning that $B_{\ell}$ obeys the explicit Laplace's determinantal formula \cite{KORN67}:

\begin{equation}
B_{\ell}=\left| \begin{array}{ccccc}
1 & 0 & \cdots & 0 & 1 \\
\frac{1}{2!} & 1 &  & 0 & 0 \\
\vdots &  & \ddots &  & \vdots \\
\frac{1}{\ell!} & \frac{1}{(\ell-1)!} &  & 1 & 0 \\
\frac{1}{(\ell+1)!} & \frac{1}{\ell!} & \cdots & \frac{1}{2!} & 0 \\
\end{array}\right|.
\end{equation}

\noindent It can be noticed that Eq. (\ref{finee2}) presents some similarities with expression (\ref{ord}). In addition, it is interesting to investigate whether Bernoulli numbers occur in known identities involving (anti-)commutators in quantum mechanics. Considering two non-commuting operators $\hX$ and $\hY$, the Baker-Campbell-Hausdorff relation mentioned above (see section \ref{sec1}) enables one to express the operator $\hZ$, defined by $\exp\left(\hX\right)\exp\left(\hY\right)=\exp\left(\hZ\right)$, as

\begin{eqnarray}
\hZ&=&\hX+\hY+\frac{1}{2}\left[\hX,\hY\right]+\frac{1}{12}\left[\hX,\left[\hX,\hY\right]\right]+\frac{1}{12}\left[\hY,\left[\hY,\hX\right]\right]\nonumber\\
& &-\frac{1}{24}\left[\hX,\left[\hY,\left[\hX,\hY\right]\right]\right]+\cdots.
\end{eqnarray}

\noindent The Hausdorff method \cite{MAGNUS54} consists in writing $\hZ$ as the summation

\begin{equation}
\hZ=\hX+\hZ_1+\hZ_2+\cdots,
\end{equation}

\noindent where $\hZ_n$ contains all the terms of degree $n$ with respect to $\hY$. As concerns the linear part, one has

\begin{eqnarray}
\hZ_1&=&\sum_{n=0}^{\infty}\frac{B_n}{n!}\left[\left[\vphantom{\hX}\right.\right.\cdots\left[\hY\right.,\underbrace{\left.\left.\hX\right]\cdots\right],\hX}_{n}\left.\vphantom{\hX}\right]\nonumber\\
&=&\sum_{n=0}^{\infty}(-1)^n\frac{B_n}{n!}\left[\vphantom{\hX}\right.\underbrace{\hX,\left[\vphantom{\hX}\right.\cdots\left[\hX\right.}_{n},\left.\hY\right]\cdots\left.\left.\vphantom{\hX}\right]\right].
\end{eqnarray}

\noindent which involves Bernoulli polynomials as well. Therefore, to first order in $\hY$, the Baker-Campbell-Hausdorff formula can be writen as \cite{KLARSFELD89}:

\begin{equation}
\hZ=\hX+\sum_{n=0}^{\infty}(-1)^n\frac{B_n}{n!}\left[\vphantom{\hX}\right.\underbrace{\hX,\left[\vphantom{\hX}\right.\cdots\left[\hX\right.}_{n},\left.\hY\right]\cdots\left.\left.\vphantom{\hX}\right]\right] +O\left(\hY^2\right).
\end{equation}

\section{\label{sec5} Quantum-classical transition, Ehrenfest theorem and time evolution of moments}

An interesting field of application involving commutators of powers of the position and momentum operators $\left[\hx^n,\hp_x^m\right]$ is the quantum-classical relationship. The Ehrenfest theorem plays a major role in the characterization of the quantum-classical transition. Writing the Hamiltonian of the system

\begin{equation}
\hH=\frac{\hp_x^2}{2m}+V\left(\hx\right),
\end{equation}

\noindent the equations of motion are

\begin{equation}
\frac{d\langle\hx\rangle}{dt}=\langle\hp_x\rangle
\end{equation}

\noindent and

\begin{equation}
\frac{d\langle\hp_x\rangle}{dt}=-\left\langle\frac{dV}{d\hx}\right\rangle.
\end{equation}

\noindent Introducing the variables $\hX=\hx-\langle\hx\rangle$ and $\hY=\hp_x-\langle\hp_x\rangle$, the Hamiltonian can be rewritten, after developping the potential $V$ in Taylor series:
 
\begin{equation}\label{hei0}
\hH=\frac{\hY^2}{2m}+\frac{\langle\hp_x\rangle\hY}{m}+\frac{\langle\hp_x\rangle^2}{2m}+\sum_{k=0}^{\infty}\frac{1}{k!}\frac{d^kV\left(\langle\hx\rangle\right)}{d\langle\hx\rangle^k}\hX^k.
\end{equation}

\noindent Therefore, the time dependence of $\hp_x$ requires the higher moments of the centered position distribution $\langle \hX^k\rangle$. In such circumstances, relations involving commutators and anti-commutators of monomials can be of interest in order to obtain compact expressions for the time evolution of these moments or their commutators (see section \ref{sec2}). Let us consider, for instance, the case of a power-law potential $V\left(\hx\right)=\hx^l$ where $l$ is an integer ($l$=2 corresponds to the harmonic oscillator, higher values of $l$ to anharmonic terms). Considering only the potential part of the Hamiltonian (i.e. omitting the kinetic energy part) and using variables $\hX$ and $\hY$, Heisenberg's equations of motion read

\begin{equation}\label{hei1}
\frac{d}{dt}\left(\hY^k\hX^n\right)=\frac{i}{\hbar}\left[\hX^l,\hY^k\hX^n\right]+\frac{\partial}{\partial t}\left(\hY^k\hX^n\right)
\end{equation}

\noindent and

\begin{equation}\label{hei2}
\frac{d}{dt}\left(\hX^n\hY^k\right)=\frac{i}{\hbar}\left[\hX^l,\hX^n\hY^k\right]+\frac{\partial}{\partial t}\left(\hX^n\hY^k\right).
\end{equation}

\noindent Combining Eqs. (\ref{hei1}) and (\ref{hei2}) enables one to write

\begin{equation}
\frac{d}{dt}\left\{\hX^n,\hY^k\right\}=\frac{i}{\hbar}\left[\hX^{n+l},\hY^k\right]-\frac{i}{\hbar}\hX^l\left[\hX^n,\hY^k\right]+\frac{i}{\hbar}\hX^n\left[\hX^l,\hY^k\right],
\end{equation}

\noindent which leads, with the use of Eq. (\ref{finee2}), to the symmetrical expression

\begin{eqnarray}
\frac{d}{dt}\left\{\hX^n,\hY^k\right\}&=&\frac{2i}{\hbar}\sum_{j=1}^{\mathrm{min}(n+l,k)}c^jj!\bin{n+l}{j}\bin{k}{j}\frac{\left(2^{j+1}-1\right)}{j+1}\nonumber\\
& &\times B_{j+1}\left\{\hX^{n+l-j},\hY^{k-j}\right\}\nonumber\\
& &-\frac{2i}{\hbar}\hX^l\sum_{m=1}^{\mathrm{min}(n,k)}c^mm!\bin{n}{m}\bin{k}{m}\frac{\left(2^{m+1}-1\right)}{m+1}\nonumber\\
& &\times B_{m+1}\left\{\hX^{n-m},\hY^{k-m}\right\}\nonumber\\
& &+\frac{2i}{\hbar}\hX^n\sum_{r=1}^{\mathrm{min}(l,k)}c^rr!\bin{l}{r}\bin{k}{r}\frac{\left(2^{r+1}-1\right)}{r+1}\nonumber\\
& &\times B_{r+1}\left\{\hX^{l-r},\hY^{k-r}\right\},\nonumber\\
\end{eqnarray}

\noindent involving only anti-commutators.

\section{\label{sec6} Conclusion}

\noindent In this short paper, we presented a relation for the commutator of two monomial operators $\hX^n$ and $\hY^m$ in the case where the commutator $\left[\hX,\hY\right]$ is a constant. The formula is an expression in terms of anti-commutators (with lower powers of $\hX$ and $\hY$) and involves Euler polynomials evaluated in zero or, equivalently, Bernoulli numbers. Generalization to the commutator of products of an arbitrary number of monomials is in progress, as well as to nests of commutators fitted into each other. It could also be interesting to extend the method presented here to $q$-deformed commutators, arising in supersymmetric theories \cite{BIEDENHARN89,DAOUD06}. We hope that such expressions will be helpul for many studies in quantum-mechanics and quantum-optics.

\vspace{5mm}
\large
\noindent {\bf Acknowledgements}
\normalsize
\vspace{5mm}

\noindent The author is indebted to Yvan Castin and Patrick Robertson for helpful suggestions. 

\appendix

\section{Appendix: Uniqueness of the solution of vector $v_k$}

Writing down explicitely Eq. (\ref{mt2}) for all the values of the index $k$, we obtain a linear system:

\begin{equation}
\left\{
\begin{array}{l}
2v_1=1\\
2v_1+2v_2=1\\
3v_1+3v_2+2v_3=1\\
4v_1+6v_2+4v_3+2v_4=1\\
5v_1+10v_2+10v_3+5v_4+2v_5=1\\
6v_1+15v_2+20v_3+15v_4+6v_5+2v_6=1\\
7v_1+21v_2+35v_3+35v_4+21v_5+7v_6+2v_7=1\\
8v_1+28v_2+56v_3+70v_4+56v_5+28v_6+8v_7+2v_8=1\\
9v_1+36v_2+84v_3+126v_4+126v_5+84v_6+36v_7+9v_8+2v_9=1\\
\;\;\;\;\;\;\;\;\vdots,
\end{array}\right.
\end{equation}

\noindent which leads to

\begin{equation}
\left\{
\begin{array}{l}
v_1=-E_1(0)=\frac{1}{2}\\
v_2=-E_2(0)=0\\
v_3=-E_3(0)=-\frac{1}{4}\\
v_4=-E_4(0)=0\\
v_5=-E_5(0)=\frac{1}{2}\\
v_6=-E_6(0)=0\\
v_7=-E_7(0)=-\frac{17}{8}\\
v_8=-E_8(0)=0\\
v_9=-E_9(0)=\frac{31}{2}\\
\;\;\;\;\;\;\;\;\vdots
\end{array}\right..
\end{equation}

\noindent and the solution $v_k=-E_k(0)$ is the unique (single) solution of the system.

\end{document}